\documentclass[twocolumn,showpacs,preprintnumbers,amsmath,amssymb,prb,aps]{revtex4-2}
\usepackage{graphicx}% Include figure files
\usepackage{bm}% bold math
\usepackage{amsmath}
\usepackage{amsfonts}
\usepackage{xcolor}
\usepackage{multirow}
\usepackage{gensymb}
\begin{document}

\title{Anomalies in the electronic, magnetic and thermal behavior near the Invar compositions of Fe-Ni alloys}

\author{Ananya Sahoo, Ayusa Aparupa Biswal, S.K. Parida and V. R. R. Medicherla}
\email{Corresponding author: mvramarao1@gmail.com}
\affiliation{Department of Physics, ITER, Siksha 'O' Anusandhan Deemed to be University, Bhubaneswar 751030, India}

\author{Soumya Shephalika Behera}
\affiliation{UGC-DAE Consortium for Scientific Research, University Campus, Khandwa Road, Indore 452001, India}

\author{M. N. Singh$^1$, A. Sagdeo$^{1,2}$}
\affiliation{$^1$ Accelerator Physics and Synchrotrons Utilization Division,
Raja Ramanna Centre for Advanced Technology, Indore 452013, India}
\affiliation {$^2$ Homi Bhabha National Institute, Training School Complex, Anushakti Nagar, Mumbai 400094, India}
\author{Sawani Datta, Abhishek Singh$^1$ and Kalobaran Maiti}
\affiliation{Department of Condensed Matter Physics and Materials Science, Tata Institute of Fundamental Research, Homi Bhabha Road, Colaba, Mumbai 400005, India}
\
\affiliation{$^1$ Present address: Department of Physics and Photonics Science, National Institute of Technology, Hamirpur, Himachal Pradesh, 177005 }

\begin{abstract}
The structural and magnetic properties of Fe$_{1-x}$Ni$_x$~($x$ = 0.32, 0.36, 0.40, 0.50) alloys have been investigated using synchrotron based x-ray diffraction (XRD) technique with x-rays of wavelength 0.63658 \AA\ down to 50 K temperature, magnetic measurement using superconducting quantum interference device (SQUID) magnetometer and high resolution x-ray photoelectron spectroscopy (XPS) with monochromatic AlK$_\alpha$ radiation. The XRD studies suggest a single phase with fcc structure for $x$ = 0.36, 0.40, and 0.50 ~alloys and a mixed phase for $x$ = 0.32 alloy containing both bcc and fcc structures. The lattice parameter of the alloys exhibits a linear dependence on temperature giving rise to a temperature independent coefficient of thermal expansion (CTE). The lowest CTE  is observed for $x$ = 0.36 Invar alloy as expected while $x$ = 0.50 alloy exhibits the highest CTE among the alloys studied. The CTE of the fcc component of mixed phase alloy is close to that of Invar alloy. The temperature dependence of magnetization of the alloys down to 2 K reveals an overall antiferromagnetic interactions within the ferromagnetic phase causing the magnetization decreasing with cooling. The field cooled and zero field cooled data show larger differences for the Invar compositions; this is also manifested in the magnetic hysteresis data at 2 K and 300 K. The Fe 2$p$ and Ni 2$p$ core level spectra exhibit spin-orbit split features along with a satellite feature in the Ni 2$p$ spectra. The spectral lineshape are almost similar for all the compositions studied. Interestingly, the spin-orbit splitting for Fe 3$p$ spectra is larger than that observed for Ni 3$p$ suggesting additional contributions due to the exchange interaction between the Fe 3$p$ core hole with the Fe 3$d$ moment. This suggests large magnetic moment contribution from Fe as expected. The core level and valence band spectra, and the magnetization data suggest significant role of disorder for the Invar compositions.
\end{abstract}

\pacs{75.50.Bb, 81.30.Bx, 61.50.-f,  79.60.—i, 71.20.-b}

\maketitle

\section{Introduction}
The physical properties of Fe-Ni system strongly depend on temperature, pressure and  composition. Fe-Ni alloys exhibit soft ferromagnetic behaviour for all compositions below the ferromagnetic transition temperature \cite{wassermann1990}. The random substitution of Ni into Fe stabilizes the $\gamma$ phase of Fe. The alloy system, Fe$_{1-x}$Ni$_x$ exhibits a structural transformation from bcc to fcc at $x=0.25$ \cite{wassermann1991,keehan,acharya2020, acharya2016}. In the intermediate composition range from $x=0.25$ to $0.30$, the alloys are of mixed phase containing both bcc and fcc structures. The striking Invar anomaly was observed in Fe-Ni alloys with Ni concentration ranging from 30-45\% \cite{abrikosov1995}. The Invar anomaly was discovered by Guillaume \cite{guillaume1904} in fcc x= 0.35 alloy in 1897 and was awarded Nobel prize in the year 1920. The Invar anomaly was attributed to moment-volume instabilities with changes in temperature \cite{weiss1963}.

The magnetism in fcc Fe$_{1-x}$Ni$_x$ alloys is complicated due to the possibility of  spectrum of magnetic states \cite{wassermann1991}. Adding further complexity, there is an evidence of anti-ferromagnetism with Fe sites pointing antiparallel to the total magnetization which leads to non-collinear magnetism in FeNi Invar alloys \cite{abrikosov2007, ruban2007}. There are so many theoretical models calculating the localized magnetic moments in Fe-Ni alloys \cite{hesse} using the exchange integrals J$_{FeFe}$, J$_{FeNi}$, and J$_{NiNi}$ as its parameters. Within this framework, first principles studies suggest that magnetic structure is highly frustrated in Fe-Ni Invar alloy. The orientation of iron magnetic moments can either be parallel or antiparallel, depending upon the local environment which offers an explanation for the magnetic anomalies observed in FeNi Invar alloys. Recently, Ehn {\it et.al} have reported ab initio study of the pressure induced Invar effect in fcc binary alloys and interestingly, they have found that both magnetic and lattice vibrational contribution cancel out each other leading to negligible thermal expansion\cite{ehn2023}. In a recent publication, the authors have calculated  components of thermal expansion arising from both magnetism and atomic vibrations in various pressure ranges \cite{lohaus2023}. The anomalous behaviour of atomic volume, elastic constants, heat capacity, spontaneous volume magnetostriction and slater pauling curve\cite{groundstate} etc. display unconventional behavior within the Invar composition range. Fe-Ni Invar alloys are utilized extensively in precision instrumentation, mechanical devices and aerospace technology \cite{ananya2021}.\\

\section{Experimental}

Polycrystalline Fe$_{1-x}$Ni$_x$ alloys have been prepared by arc melting technique as explained in detail in the previous paper \cite{ananya2024}. X-ray powder diffraction (PXRD) was carried out at temperatures ranging from 50~K to 300~K using Angle dispersive x-ray diffraction (ADXRD) beamline (BL-12) at Indus-2 synchrotron source, Raja Ramanna Centre for Advanced Technology (RRCAT), Indore, India. The PXRD profiles were collected with a MAR345 image plate detector, and the resulting two dimensional patterns were integrated into 2q scales with the FIT2D software. The incident X-ray wavelength was calibrated using the data from a reference LaB$_6$ sample, and the estimated value is found to be 0.63658 \AA. Magnetization measurement on Fe$_{1-x}$Ni$_{x}$ alloys has been performed using superconducting quantum interference device vibrating sample magnetometer (SQUID VSM) system. X-ray photo electron spectroscopy (XPS) measurements were carried out at TIFR, Mumbai using a Phoibos150 analyzer ($\Delta{E}$ = 0.35 eV) from Specs GmbH at room temperature with monochromatic Al K$_{\alpha}$ (1486.6 eV) x-ray source. Samples were scrapped in high vacuum to obtain contaminant free surfaces of the alloys. The photoelectron spectra were recorded at a base pressure of 10$^{-10}$ Torr. Core level data fitting has been performed with Doniach-sunjic line shape using CASAXPS software \cite{doniach}. Binding energy of the photoelectrons was calibrated using the Fermi step taken on clean silver surface. 

\section{Results and discussions}

\begin{figure}[]
\centering{\includegraphics[scale=0.4]{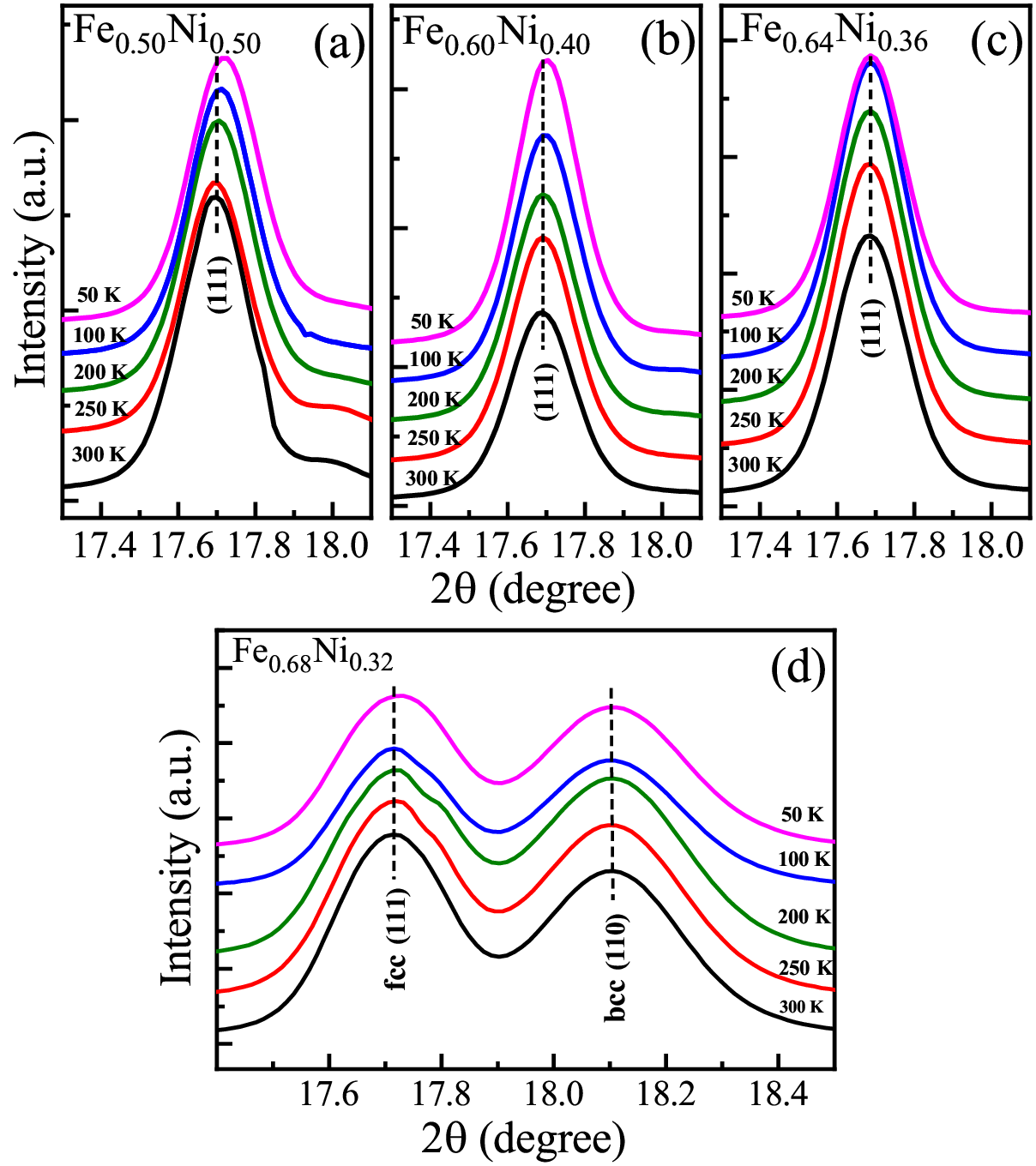}}
\caption{Synchrotron x-ray diffraction patterns of Fe$_{1-x}$Ni$_{x}$ alloys for $x$ = (a) 0.50, (b) 0.40, (c) 0.36 and (d) 0.32 at different temperatures as indicated in the figure. The x-ray wavelength used is 0.63658 \AA. The data for Fe$_{0.68}$Ni$_{0.32}$ sample show two broad peaks corresponding to fcc and bcc phases.}
\label{fig:1}
\end{figure}

The XRD patterns recorded on Fe$_{1-x}$Ni$_{x}$ alloys using synchrotron radiation of wavelength of 0.63658 \AA in a temperature range of 50 K to 300 K indicate single phase fcc crystal structure for $x$ = 0.36, 0.40 ~and~ 0.50 alloys. The XRD pattern of $x$ = 0.32 alloy exhibits two broad peaks at about $2\theta$=17.7$^\circ$ and 18.1$^\circ$ corresponding to fcc (111) and bcc (110) Bragg reflections indicating a mixed phase alloy. These XRD patterns are in agreement with the patterns recorded reported earlier using Cu K$\alpha$ radiation on these alloys \cite{ananya2024}. The (111) Bragg peak of $x$ = 0.36, 0.40 and 0.50 alloys at various temperatures is shown in Figs. \ref{fig:1}(a)-(c). For the mixed phase alloy ($x$ = 0.32), both (111) and (110) Bragg peaks are shown in Fig. \ref{fig:1}(d). In Fe$_{0.50}$Ni$_{0.50}$ alloy, a gradual shift of (111) Bragg peak to higher 2$\theta$ is observed as the temperature is lowered from 300 K to 50 K. In the case of Fe$_{0.60}$Ni$_{0.40}$ alloy, the peak-shift with the temperature is  insignificant except at 50 K. In Fe$_{0.64}$Ni$_{0.36}$ Invar alloy, there is no noticeable shift in Bragg peak with the temperature suggesting negligible change in structural parameters with temperature as expected. In the mixed phase alloy, the shift of both (111) and (110) Bragg peaks with temperature is small compared to that observed in Fe$_{0.50}$Ni$_{0.50}$ alloy.

\begin{figure}[]
\centering{\includegraphics[scale=0.4]{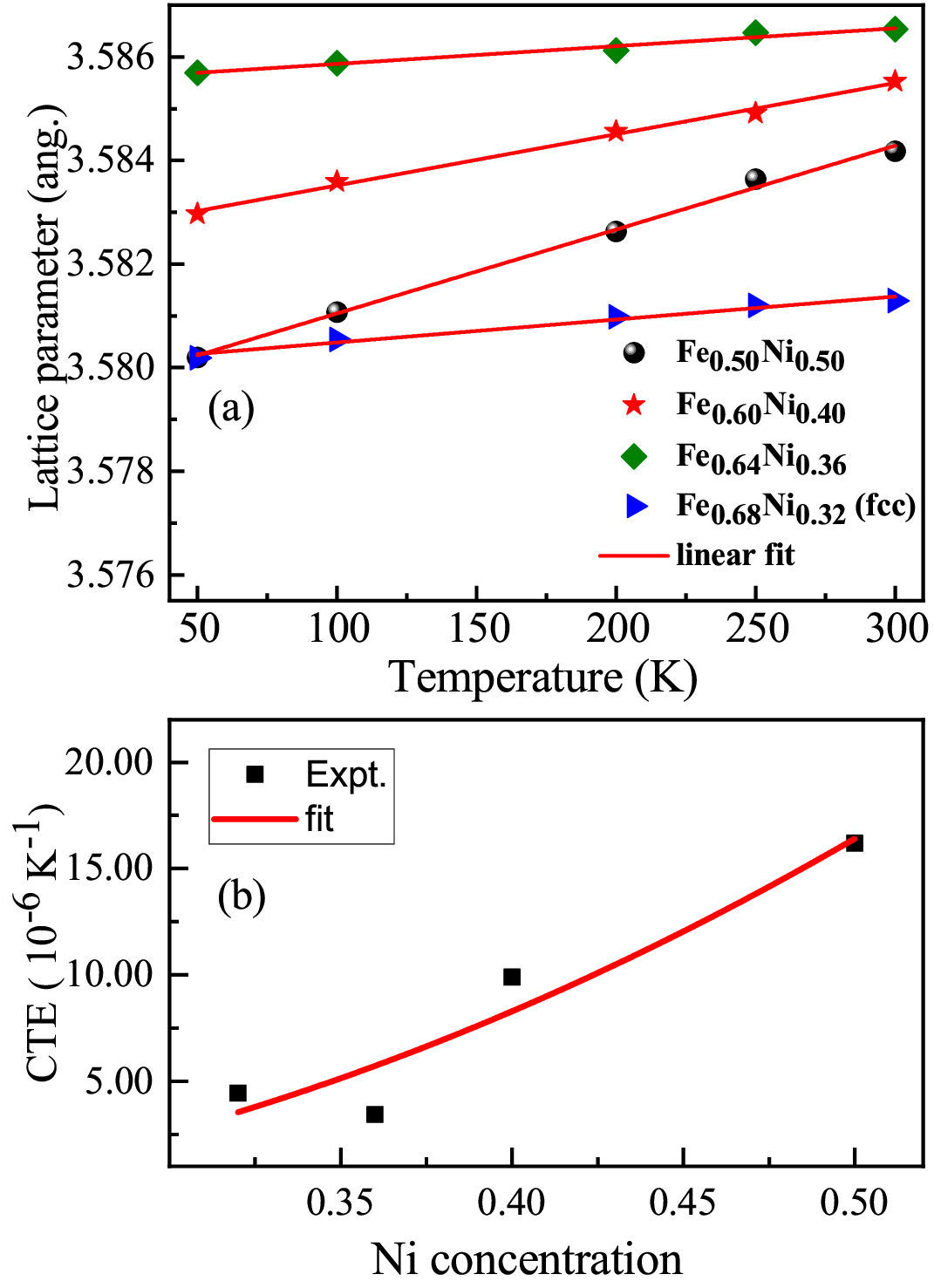}}
\caption{(a) The symbol represents the temperature dependence of lattice parameter for Fe$_{1-x}$Ni$_{x}$  alloys. The line represents the linear fit to the data. (b) The  filled squares represent the variation of CTE with Ni concentration. The line represents the numerical fit to the data using a second order polynomial}
\label{fig:2}
\end{figure}

The temperature dependence of lattice parameter for all compositions of the alloy system is shown in Fig. \ref{fig:2} (a). In case of mixed phase alloy, the lattice parameter of fcc component is plotted. The lattice parameter of each alloy is numerically fitted to a straight line as a function of temperature as shown in the Figure. The slope of the line provides the coefficient of thermal expansion, CTE. It is observed from Fig. \ref{fig:2} (a) that the temperature dependence of lattice parameter of both Invar alloy and the mixed phase alloy are almost parallel indicating similar CTE. Fig. \ref{fig:2} (b) shows the dependence of CTE on $x$, the Ni concentration. The experimental CTE is fit to a second order polynomial in x and shown as continuous line in the figure. Evidently, the lowest CTE ($3.43\times10^{-6}K^{-1}$) is observed for the Invar alloy ($x$ = 0.36) while the highest CTE ($1.6\times10^{-5}K^{-1}$) for $x$ = 0.50 alloy. The CTE exhibits an increasing trend from $x$ = 0.36 to $x$ = 0.50. The observed CTE is in agreement with previous reports \cite{guillaume1967, wegener2021}.

\begin{figure}[]
\centering{\includegraphics[scale=0.4]{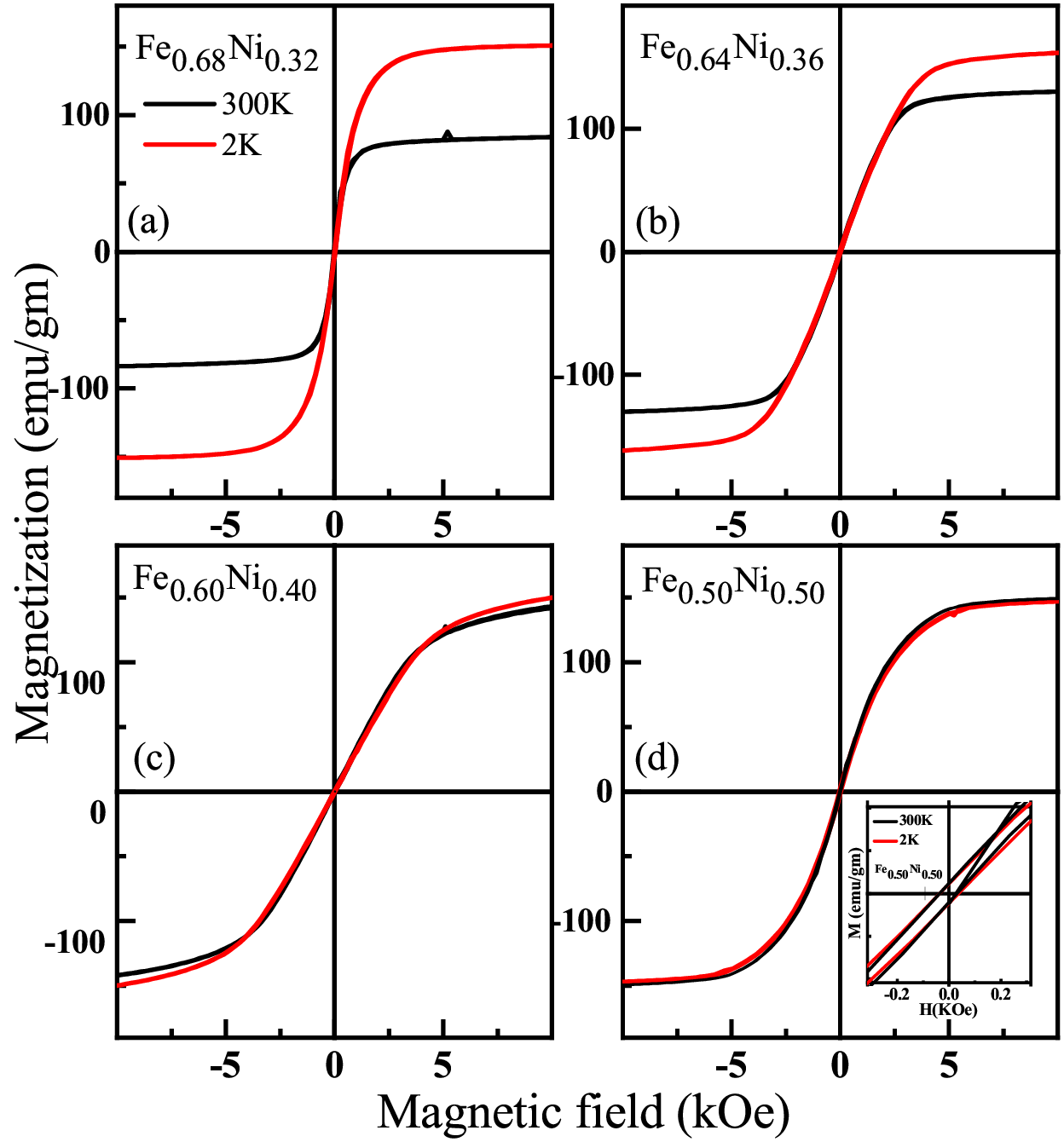}}
\caption{M-H hysteresis loops of Fe$_{1-x}$Ni$_{x}$ alloys for $x$ = (a) 0.32, (b) 0.36, (c) 0.40 and (d) 0.50 measured at 300 K (black lines) and 2 K (red lines). The inset in (d) provides an enlarged view of the hysteresis loop for $x$ = 0.50 alloy.}
\label{fig:3}
\end{figure}

Magnetization versus applied magnetic field of Fe$_{1-x}$Ni$_{x}$ alloys at the temperatures, 300 K and 2 K are shown in Fig. \ref{fig:3} (a) to (d). The hysteresis loops show very low coercive fields (10 - 30 Oe) with small remanent magnetizations at both the temperatures studied (see the inset of Fig. \ref{fig:3}(d)), indicating soft ferromagnetic behaviour \cite{acharya2021, sharmamagneto2022}. The saturation magnetization, $M_s$ at 300 K increases with the increase in Ni concentration in the alloys as evident from the M$-$H loops. However, at 2 K, the M$_s$ is almost independent of the composition. The magnetic moments of Fe  and Ni in the respective pure metals are about 2.2 $\mu_B$ and 0.6 $\mu_B$ respectively \cite{crangle1963}. Interestingly, the saturation moment in $x$ = 0.50 sample is identical at 2 K and 300 K; no temperature dependence in this wide temperature range while it is very different for $x$ = 0.32 sample.

In Fe-Ni alloys, the major contribution to magnetic moment comes from the number of unpaired 3$d$ electrons in Fe, which depends on the composition of the alloy \cite{nilesh2015, vitta2008}. The experimental results shown in Fig. \ref{fig:3} exhibit a significantly different behavior; despite reduction of Fe-content in the alloy, the saturation moment at 2 K remains almost unchanged while it is much lower at 300 K for higher Fe-concentrations. This suggests that at $x$ = 0.50, Fe contains the highest fraction of unpaired 3$d$ electrons which gradually reduces with decrease in Ni-concentration. The other striking result is that M$_s$ exhibits a strong dependence on temperature at low Ni concentrations, whereas at high Ni concentrations ($x$ = 0.40, 0.50), M$_s$ is almost independent of temperature. These observations suggest a strong coupling at high Ni concentrations ($x$ = 0.40, 0.50). The coupling in low Ni concentration alloys ($x$ = 0.32, 0.36) is weak which could be affected significantly by the thermal energy leading to low saturation magnetization at room temperature. Thus, excitation of magnetic degrees of freedom is possible for the Invar compositions while it is not so evident at higher Ni-concentrations.

\begin{figure}[h!]
\centering{\includegraphics[scale=0.4]{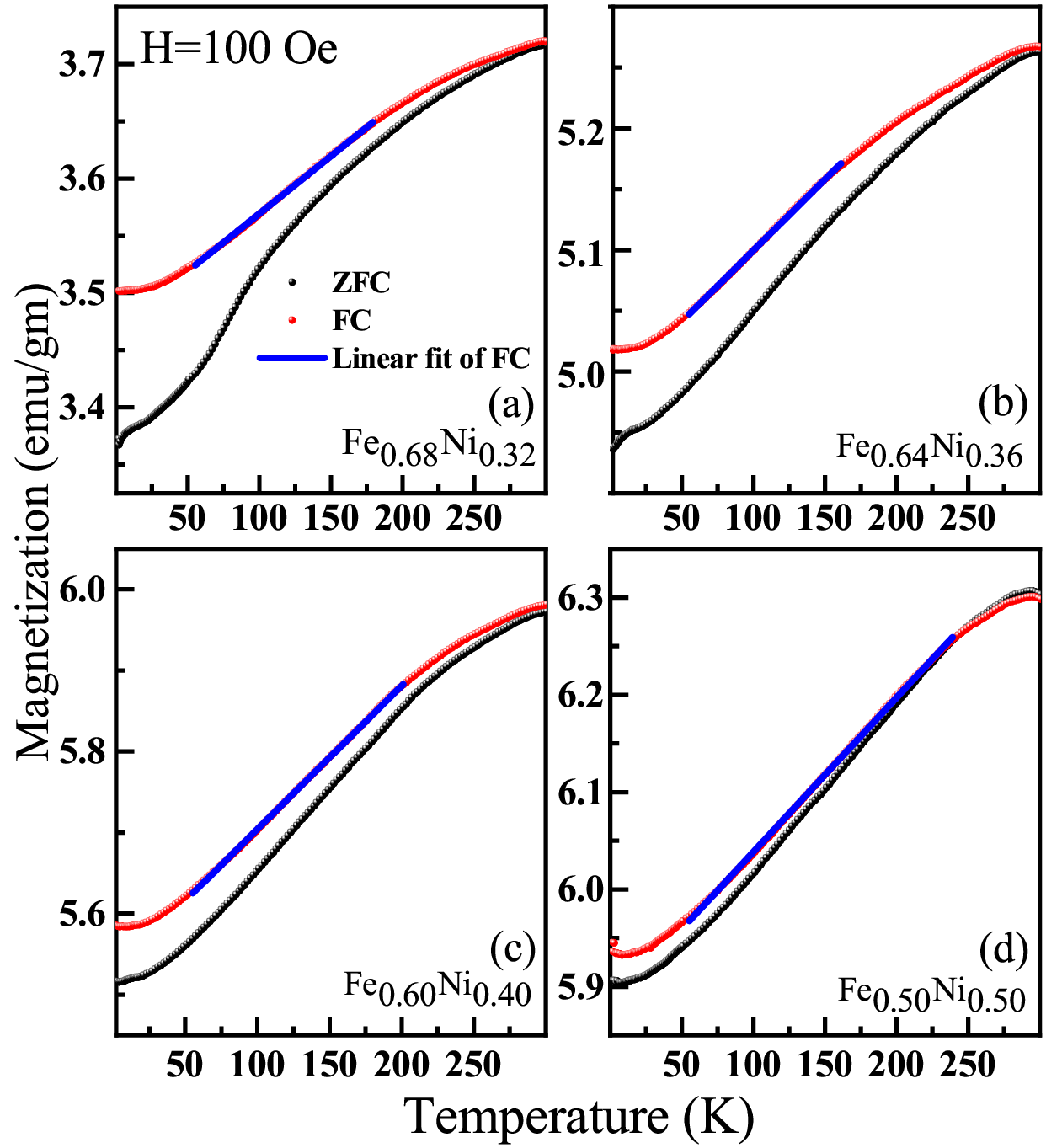}}
\caption{ZFC (black dots) and FC (red dots) magnetization curves for Fe$_{1-x}$Ni$_{x}$ alloys for $x$ = (a) 0.32, (b) 0.36, (c) 0.40, and (d) 0.50 measured at an applied magnetic field of H = 100 Oe. The blue lines in each case show the linear region.}
\label{fig:4}
\end{figure}

In Fig. \ref{fig:4}(a)-(d), we show the temperature-dependent magnetization of Fe$_{1-x}$Ni$_x$ alloys, measured from 2 to 300 K under zero-field-cooled (ZFC) and field-cooled (FC) conditions at an applied magnetic field of 100 Oe. Both ZFC and FC curves exhibit a decrease in magnetization when temperature is reduced from 300 K and the FC curves of all alloys and ZFC curves of $x$ = 0.40, 0.50 show saturation at the lowest (2 K) and highest (300 K) temperatures. Since the magnetic moment is not expected to reduce with decrease in temperature where no change in crystal field is observed, the decrease in moment may be attributed to an overall antiferromagnetism (AFM). The saturation around 300 K may be due to the ordering temperature in proximity of 300 K. The FC data taken at 2 K with 100 Oe field leads to higher magnetic moments as expected. We observe that the difference between FC and ZFC data becomes smaller with increase in Ni substitution due to stronger AFM coupling at higher Ni concentrations.

The magnetization data for the mixed phase and the Invar alloys exhibit a downward trend below 10 K indicating further reduction in magnetic moment. The FC curves of all the alloys studied exhibit a linear region above 50 K. The linear region in FC curve of each alloy is numerically fit to a straight line and shown as blue colored line in the figure. The upper temperature limit of linearity in FC curves depends on the composition of the alloy.  For $x$ = 0.32, 0.36, 0.40, 0.50 alloys, the upper temperature limits of linearity are 180 K, 160 K, 200 K and 240 K respectively. The Invar alloy ($x$ = 0.36) has the narrowest linear region.

The above discussed magnetization data as a function of temperature suggests manifestation of AFM within the ferromagnetic phase. The Fe-Ni alloys in the Invar region, possess some AFM domains \cite{kondorskii1959, kondorsky1960, colling1970}. M$\ddot{o}$ssbauer studies also suggested the existence of AFM in Fe-Ni alloys \cite{hesse, abd}. The Fe-Ni Invars  possess mixed-type exchange interactions with Fe-Fe coupling anti-ferromagnetically whereas Fe-Ni \& Ni-Ni coupling ferromagnetically \cite{rancourt1989}. When the temperature is lowered, AFM domains start to increase in number and lead to decrease the magnetization down to about 20 K at which AFM domains fully develop and hence saturation is seen in FC measurements. The downward trend observed in mixed phase and Invar alloy in ZFC measurement may be related to the formation of a new magnetic phase such as spin glass \cite{wassermann1990}.

\begin{figure}[]
\centering{\includegraphics[scale=0.4]{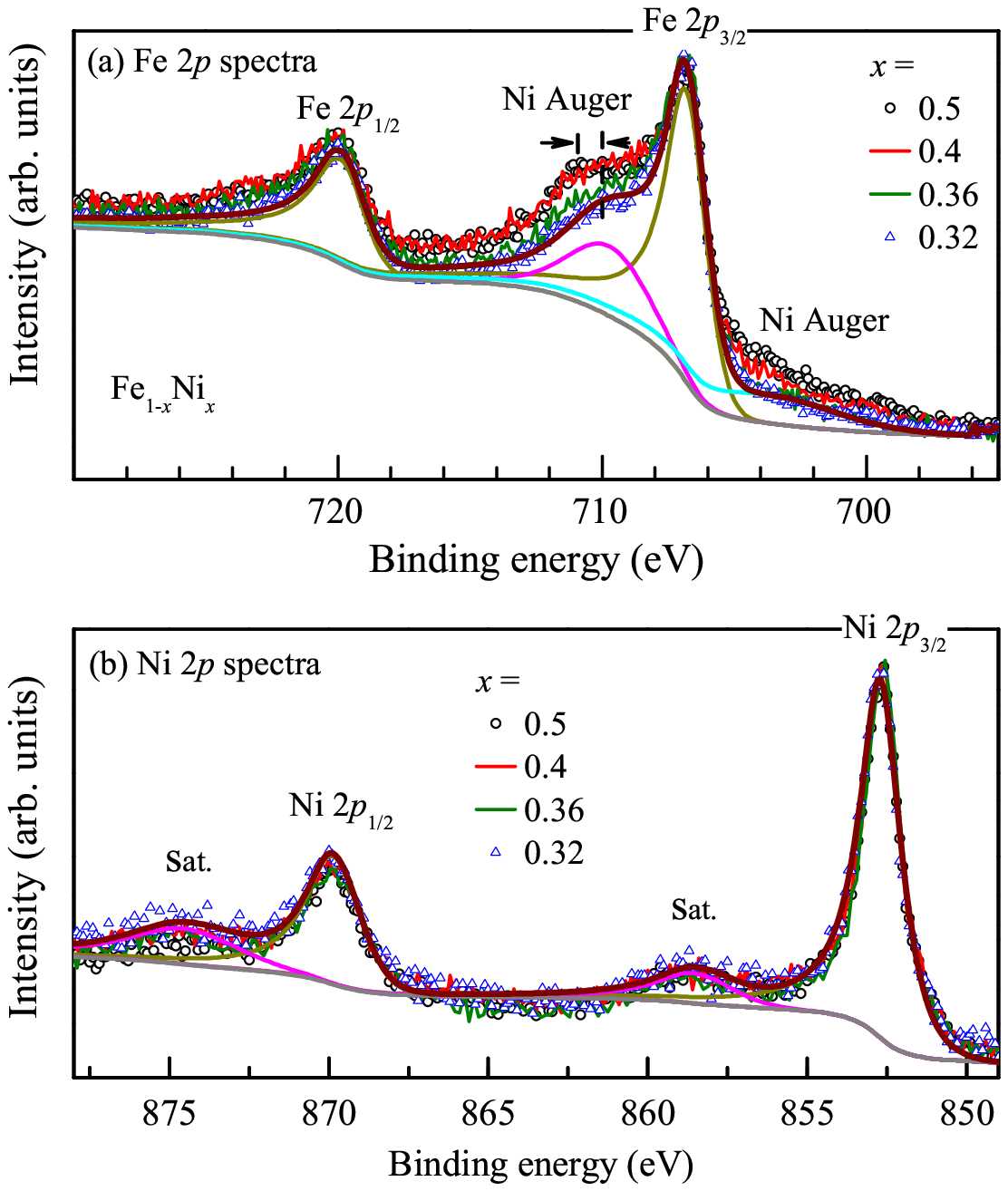}}
\caption{(a) Fe 2$p$ and (b) Ni 2$p$ spectra recorded at room temperature for Fe$_{1-x}$Ni$_x$ alloys with  $x$ = 0.50, 0.40, 0.36 and 0.32. Lines at the bottom are the fit of the experimental spectrum for $x$ = 0.32 using least square error methods and a set of DS peaks.}
\label{fig:G2pCore}
\end{figure}

We now turn to the investigation of the electronic structure which might provide some underlying scenario for the puzzles discussed above. Fe 2$p$ spectra of Fe$_{1-x}$Ni$_{x}$ alloys ($x$ = 0.32, 0.36, 0.40 and 0.50) are shown in Fig. \ref{fig:G2pCore}(a). The experimental data exhibit two sharp features corresponding to spin-orbit split 2$p_{3/2}$ and 2$p_{1/2}$ photoemission signals at binding energies 706.9~eV and 720~eV respectively with a spin orbit splitting of about 13.1 eV. In addition, we observe broad intensities between 700 - 705 eV and an intense peak at about 710.5 eV binding energies due to Ni $L_3M_{23}M_{23}$ Auger signal as also evidenced in earlier studies \cite{acharya2021}. The intensity of these features become stronger with the increase in Ni-concentration as expected. Apart from this change, we do not observe significant spectral changes with Ni-substitution.

The spectra are fitted using a set of DS function and considering Shirely background function - the results for $x$ = 0.32 case is shown in the lower panel of the Fig. \ref{fig:G2pCore}(a) by lines. The experimental results could be captured well considering single peak structure for the Fe 2$p_{3/2}$ and 2$p_{1/2}$ signals and the features for Ni Auger signals. We do not observe any change in lineshape of the Fe 2$p$ signal with Ni-substitutions though the disorder is expected to enhance with Ni-concentration. Ni Auger signal appear to shift towards higher binding energies with the increase in Ni-concentration: 710 eV for $x$ = 0.32 and 710.5 eV for $x$ = 0.50.

Experimental Ni 2$p$ spectra are shown in Fig. \ref{fig:G2pCore}(b) for all the compositions exhibiting almost identical spectral functions within the experimental noise for all the compositions. The spin-orbit split Ni 2$p_{3/2}$ and Ni 2$p_{1/2}$ signals are observed at 852.7 eV and 869.9 eV binding energies respectively suggesting a spin orbit splitting of about 17.2 eV. In addition, we observe distinct satellite intensities at about 6 eV higher binding energy. Photoemission creates a hole in the occupied level. In the presence of electron correlation between the core hole and valence electrons, the photoemission final states will be different from the ground state Hamiltonian leading to multiple features representing different screening levels of the photohole. The intense peak at lower binding energy is the well-screened feature where the core-hole is screened by the valence electrons. The feature at higher binding energy is called, satellite which corresponds to the 'poorly screened' final state; the core hole is not screened. While 3$d$ electrons in both Fe and Ni are strongly correlated, the screening appears to be strong in the case of Fe compared to Ni. This is not unusual as the exchange split Fe 3$d$ downspin band will be essentially empty to allow hopping of electrons from the conduction band to the 3$d$ level for core hole screening. The occupancy of Ni 3$d$ downspin band will be close to fully filled configuration. The intensity of the Ni 2$p$ satellite appears to remain almost the same in all the compositions studied. These results emphasize the fact that the magnetic moment of the system is primarily contributed by Fe.

\begin{figure}[]
\centering{\includegraphics[scale=0.4]{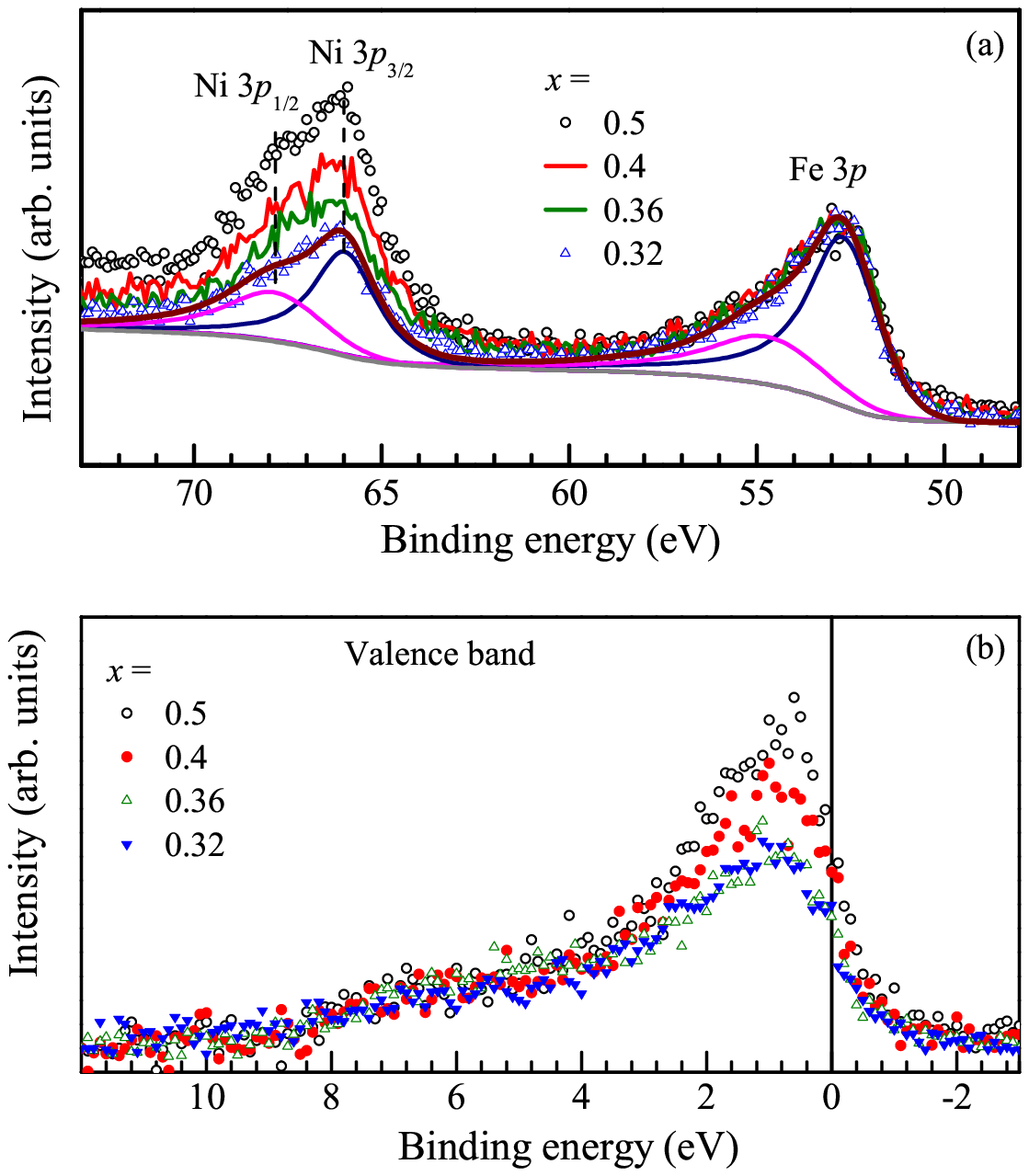}}
\caption{(a) Fe 3$p$ and Ni 3$p$ spectra of Fe$_{1-x}$Ni$_{x}$ alloys for $x$ = 0.50 (open circles), 0.40 (red line), 0.36 (green line) and 0.32 (open triangle) at 300 K. The lines at the bottom represent the least square error fit of the data for $x$ = 0.32. All the data are normalized by the intensity of the Fe 3$p$ peak. (b) Valence band spectra for all the compositions. Data are normalized by the intensity at around 6 eV.}
\label{fig:GVb3p}
\end{figure}

In Fig. \ref{fig:GVb3p}(a), we show the 3$p$ core level spectra of Fe and Ni collected at room temperature. All the spectra are normalized by the intensity of the Fe 3$p$ signal. We observe gradual increase in Ni 3$p$ spectral intensities with the increase in Ni concentration as expected. Both the core level spectra exhibit two peak structures. The main peak of Fe 3$p$ core level spectra appears at 52.8 eV binding energy along with a shoulder at about 55 eV suggesting a splitting of about 2.2 eV. Ni 3$p$ spectra show two peak structure with the peaks at 66 eV and 68 eV exhibiting a splitting of about 2 eV. None of the spectra show significant change in lineshape with the change in composition. The splitting of the 3$p$ signal can be attributed to the spin-orbit coupling though there may be contributions from the exchange interaction between the 3$p$ hole and valence electrons \cite{ananya2024}. It is to note here that although Ni is relatively heavier and the Ni 3$p$ binding energy is higher than Fe 3$p$, the splitting of Fe 3$p$ signal is found to be somewhat larger indicating additional contributions. This suggests finite contribution from the exchange splitting; this contribution will be less significant for Ni as Ni 3$d$ downspin band is close to the filled configuration \cite{chanderesis}.

The valence band spectra of all the samples are shown in Fig. \ref{fig:GVb3p}(b); the data are normalized by the intensity around 6 eV binding energy after subtracting Shirley background. The spectra look almost identical for $x$ = 0.32 and 0.36. For higher Ni concentrations, the intensity near the Fermi level starts increasing gradually. Since the occupancy of Ni 3$d$ is larger than Fe 3$d$, enhancement of intensity near the Fermi level is expected as observed at higher Ni concentrations. However, absence of such enhancement for the $x$ = 0.36 sample suggests influence of disorder \cite{EuFe2As2} which often leads to depletion of intensity close to the Fermi level.

\section{conclusion}

In summary, we studied the electronic properties of Fe$_{1-x}$Ni$_x$~($x$ = 0.32, 0.36, 0.40, 0.50) using magnetization and electron spectroscopy techniques. The samples have been prepared by arc melting method and the structural, magnetic and electronic properties were studied. The crystal structure of the alloys is investigated  using x-ray diffraction (XRD) technique employing  synchrotron radiation of wavelength 0.63658 \AA\  down to 50~K temperature exhibiting weak temperature dependence of the structural parameter for higher Ni-concentrations. The Invar compositions do not show temperature dependence. The coefficient of thermal expansion (CTE) in the alloys is independent of temperature. The lowest CTE (3.43$\times$10$^{-6}$ K$^{-1}$ ) is observed for $x$ = 0.36 Invar alloy and $x$ = 0.50 alloy exhibits the highest CTE (1.6$\times$10$^{-5}$ K$^{-1}$) among the alloys studied.

The magnetic properties of the alloys are studied down to 2 K temperature using a SQUID magnetometer exhibiting interesting behavior. While all the samples show soft ferromagnetic behavior, the temperature dependent magnetization curve reveal overall antiferromagnetic behavior. The effect of temperature on saturation moment is almost negligible at higher Ni-concentration while it is strong in the Invar compositions.

The core levels and the valence bands of the alloys were investigated using high resolution x-ray photoelectron spectroscopy employing Al K$_\alpha$ source. Fe 2$p$ and Ni 2$p$ spectra appear almost similar in all the compositions; Ni 2$p$ spectra show signature of electron correlation induced satellites. The spin-orbit splitting in Fe 3$p$ appears to be larger than Ni 3$p$ spectra suggesting additional contributions presumably due to interaction of Fe 3$p$ core hole spin with the Fe 3$d$ moment which will be suppressed in the case on Ni. The spin-orbit splitting trend in the 2$p$ photoemission is in the expected line. The valence band spectra is found to be very similar for $x$ = 0.32 and 0.36; the intensity near the Fermi level starts increasing for higher Ni concentrations. This suggests important role of disorder in the Invar compositions as also manifested in the magnetization data.\\

\section*{Acknowledgment}

Authors acknowledge the financial support from the Department of Atomic Energy (DAE), Govt. of India (Project Identification no. RTI4003, DAE OM no. 1303/2/2019/R\&D-II/DAE/2079 dated 11.02.2020). K. M. acknowledges financial support from BRNS, DAE, Govt. of India under the DAE-SRC-OI Award (grant no. 21/08/2015-BRNS/10977)  and J. C. Bose Fellowship (SB/S2/JCB-24/2014), DST, Govt. of India.\\

The authors AS and VRRM acknowledge Dr. Rajeev Rawat, UGC-DAE Consortium for Scientific Research, Indore, India for extending sample preparation facilities for this work.

%\pagebreak


\begin{thebibliography}{2}

\bibitem{wassermann1990}
E. F. Wasserman,  Ferromagnetic Materials, Vol.{\bf 5} (eds  K. H. J. Buschow and E. P. Wohlfarth), (North-Holland, Amsterdam, 1990) 237-322.

\bibitem{wassermann1991}
E. F. Wassermann, J. Magn. Magn. Mater. {\bf 100} (1991) 346–362.

\bibitem{keehan}
L.W. McKeehan, Phys. Rev. {\bf 21}
(1923) 402–407.

\bibitem{acharya2020}
Acharya, S. S., et al. Journal of Electron Spectroscopy and Related Phenomena {\bf 240} (2020) 146933.

\bibitem{acharya2016}
S.S. Acharya, V.R.R. Medicherla, R. Rawat, K. Bapna, K. Ali, D. Biswas, K. Maiti, J. Electron Spectrosc. Relat. Phenom. {\bf 212} (2016) 1–4.

\bibitem{abrikosov1995}
Abrikosov, I. A., Eriksson, O., S$\ddot{o}$derlind, P., Skriver, H. L., \& Johansson, B. Physical Review B {\bf 51}(1995) 1058.

\bibitem{guillaume1904}
GUILLAUME, C. Nature {\bf 71} (1904) 134–139.

\bibitem{weiss1963}
Weiss, R. J. Proc. Phys. Soc. {\bf 82} (1963) 281.

\bibitem{abrikosov2007}
Abrikosov, I. A. et al. Phys. Rev. B {\bf 76} (2007) 014434.

\bibitem{ruban2007}
Ruban, Andrei V., et al. Physical Review B {\bf 76} (2007) 014420.

\bibitem{hesse}
(a) J. Hesse, Z. Phys. B {\bf 54} (1983) 35–42, https://doi.org/10.1007/ BF01507947; \\
(b) M. Shiga and Y. Nakamura, J. Magn. Magn. Mater. {\bf 40} (1984) 319–327, https://doi. org/10.1016/0304-8853(84)90324-X; \\ 
(c) H. Ullrich and J. Hesse, J. Magn. Magn. Mater. {\bf 45} (1984) 315–327, https://doi.org/ 10.1016/0304-8853(84)90025-8; \\
(d) V. A. Makarov, I. M. Puzei, T. V. Sakharov, and O. V. Basar-Sov. Phys. {\bf 61}  (1985) 839.

\bibitem{ehn2023}
Ehn, A., Alling, B., \& Abrikosov, I. A. Physical Review B {\bf 107} (2023) 104422.

\bibitem{lohaus2023}
Lohaus, S.H., Heine, M., Guzman, P. et al. Nat. Phys. {\bf 19} (2023) 1642–1648 .

\bibitem{groundstate}
(a)  V. L. Sedov, JEP Lett. {\bf 14} (1971) 341; \\
(b) G. Blaise and M. C. Cadeville, J. Phys. (Paris) {\bf 36} (1975) 545–550 ;\\
(c) S. F. Dubinin, S. G. Teplouchov, S. K. Sidorov, Y. U. Izyumov, and V. N. Syromyatnikov, Phys. Stat. Solidi A {\bf 61} (1980) 159–167 ;\\
(d) Yoji Nakamura, Yasuo Takeda, and Masayuki Shiga, J. Phys. Soc. Jpn. {\bf 25} (1968) 287.

\bibitem{ananya2021}
Sahoo, A., and V. R. R. Medicherla. "Fe-Ni Invar alloys: a review." Materials today: proceedings {\bf 43} (2021) 2242-2244.

\bibitem{ananya2024}
Sahoo, A., and V. R. R. Medicherla. Journal of Alloys and Compounds {\bf 994} (2024) 174544

\bibitem{doniach}
S Doniach and M Sunjic J. Phys. C: Solid State Phys.   {\bf 3} (1970) 285.

\bibitem{guillaume1967}
C. E. Guillaume, Nobel Lectures, Physics 1901-1921 (Elsevier, Amsterdam,
1967).

\bibitem{wegener2021}
Wegener, Thomas, et al. 
%"On the structural integrity of Fe-36Ni Invar alloy processed by selective laser melting." 
Additive Manufacturing {\bf 37} (2021) 101603.

\bibitem{acharya2021}
Acharya, SS and Medicherla, VRR and Bapna, Komal and Ali, Khadiza and Biswas, Deepnarayan and Rawat, Rajeev and Maiti, Kalobaran, Journal of Alloys and Compounds {\bf 863} (2021) 158605.

\bibitem{sharmamagneto2022}
Sharma, Mohit K and Kumar, Akshay and Kumari, Kavita and Park, Su-Jeong and Yadav, Naveen and Huh, Seok-Hwan and Koo, Bon-Heun. Magnetochemistry {\bf 9} (2022) 8.

\bibitem{crangle1963}
J. Crangle and G. C. Hallam,
%The Magnetization of Face-Centred Cubic and Body-Centred Cubic Iron+Nickel Alloys, 
Proc. R. Soc. Lond. A {\bf 272} (1963) 119.

\bibitem{nilesh2015}
N.S Kanhe, A. Kumar, S.M. Yusuf, A.B. Nawale, S.S. Gaikwad, S.A. Raut,
S.V. Bhoraskar, S.Y. Wu, A.K. Das, V.L. Mathe, Journal of Alloys and Compounds {\bf 663} (2015) 30-40.

\bibitem{vitta2008}
S. Vitta, A. Khuntia, G. Ravikumar, D. Bahadur, Journal of magnetism and magnetic materials {\bf 320} (2008) 182-189.

\bibitem{kondorskii1959}
E.I. Kondorskii, JETP 10 (1960) 1284; J. Expt. Theoret.
Phys. (USSR) {\bf 37} (1959) 1819.

\bibitem{kondorsky1960}
E.I. Kondorsky and V.L. Sedov, J. Appl. Phys. {\bf 31} (1960) 331.

\bibitem{colling1970}
D.A. Colling and W.J. Carr Jr., J. Appl. Phys. {\bf 41} (1970) 5125.

\bibitem{abd}
M.M. Abd-Elmeguid, U. Hobuss, H. Micklitz, B. Huck, J. Hesse,
Phys. Rev. B {\bf 35} (1987) 4796–4800.

\bibitem{rancourt1989}
Rancourt, D. G., S. Chehab, and G. Lamarche. Journal of magnetism and magnetic materials {\bf 78} (1989) 129-152.

\bibitem{chanderesis}
Chandesris, D., J. Lecante, and Y. Petroff. Phys. Rev. B {\bf 27} (1983) 2630.

\bibitem{EuFe2As2}
G. Adhikary, N. Sahadev, D. Biswas, R. Bindu, N. Kumar, A. Thamizhavel, S. K. Dhar, and K. Maiti,
J. Phys.: Condens. Matter {\bf 25} (2013) 225701.

%\bibliographystyle{plainnat}
%\bibliography{mybibfile}
\end{thebibliography}
\end{document}